\documentclass[11pt]{article}

\usepackage[final]{acl}

\usepackage{times}
\usepackage{latexsym}

\usepackage[T1]{fontenc}
\usepackage{placeins}

\usepackage[utf8]{inputenc}

\usepackage{inconsolata}

\usepackage{algorithm}
\usepackage{algpseudocode}
\usepackage{float}
\usepackage{microtype}
\usepackage{pgfplots}
\pgfplotsset{compat=1.18}
\usepackage{tikz}
\usepackage[skins,breakable,listings]{tcolorbox}
\usepackage{amsmath,amssymb,amsfonts}
\usepackage{amsthm}
\usepackage{enumitem}
\usepackage{stmaryrd}
\usepackage{graphicx}
\usepackage{subcaption}
\usepackage{booktabs}
\usepackage{tabularx}
\usepackage{multirow}
\usepackage{xcolor}
\usepackage{hyperref}
\usepackage[capitalise,noabbrev]{cleveref}
\usepackage{siunitx}
\newcommand{\method}{\textsc{Reflect-SQL}} %
\newcommand{\datasetSpider}{Spider 1.0}
\newcommand{\datasetBird}{BIRD}
\usepackage{listings}

\newcolumntype{Q}{>{\ttfamily\footnotesize\arraybackslash}p{0.55\linewidth}}
\newcolumntype{R}{>{\raggedright\arraybackslash}p{0.35\linewidth}}
\title{Reflective Reasoning for SQL Generation}

\author{
 \textbf{Isabelle Mohr\textsuperscript{1,2}},
 \textbf{Joao Pedro Gandarela\textsuperscript{1}},
 \textbf{John Dujany\textsuperscript{2}},
 \textbf{Andre Freitas\textsuperscript{1}},
\\
 \textsuperscript{1}Idiap Research Institute,
 \textsuperscript{2}Merck KGaA,
\\
 \small{
   \textbf{Correspondence:} \href{mailto:isabelle.mohr@idiap.ch}{isabelle.mohr@idiap.ch}
 }
}

\begin{document}
\maketitle
\begin{abstract}

Robust text-to-SQL over complex, real-world databases remains brittle even with modern LLMs: iterative refinement often introduces syntactic and semantic drift, corrections tend to be non-transferable across queries, and naive use of large context windows scales poorly. We propose a controlled text-to-SQL framework built around \emph{reflective refinement}. Instead of repeatedly rewriting the current SQL instance, the system decomposes generation into typed stages and applies feedback as persistent updates to the \emph{stage-level generation mechanism}. A Reflection-Refinement Loop localizes violations to the responsible stage maximize preservation of previously validated constraints and support monotonic improvement over a query set. The method operates without gold SQL by combining interpreter-based checks with LLM-based semantic coverage verification as epistemic judges. Experiments on Spider and BIRD demonstrate consistent gains over strong prompting baselines, robust convergence within a small refinement budget, and improved execution accuracy across both frontier and open-weight model families.
\end{abstract}


\section{Introduction}

While Large language models (LLMs) have substantially advanced natural-language interfaces to databases, still robust text-to-SQL generation over complex/real-world databases remains unsolved \citep{yu2019spiderlargescalehumanlabeleddataset,wang2021ratsqlrelationawareschemaencoding}. 

Recent work has increasingly shifted from single-shot semantic parsing toward \emph{LLM-driven, multi-stage} text-to-SQL pipelines that decompose the task and incorporate iterative correction. Decomposition-based prompting and self-correction improve robustness by separating schema interpretation, predicate grounding, and SQL construction \citep{DBLP:conf/nips/PourrezaR23}, while execution-guided refinement strengthens reliability by using runtime signals to diagnose and repair errors \citep{mao-etal-2024-enhancing}. In parallel, agentic reasoning patterns that interleave reasoning with feedback have popularized iterative refinement as a general strategy \citep{yao2023reactsynergizingreasoningacting}. However, these approaches typically apply refinement by \emph{rewriting the specific SQL instance}, often via partial or full regeneration, rather than improving the underlying stage prompts/mechanisms, so corrections can be brittle and non-transferable across queries \citep{wang2023selfconsistencyimproveschainthought}. 

\paragraph{The Phenomena: Why is text-to-SQL an intrinsically a hard problem?} 
\medskip
\paragraph{(Challenge 1)}\textit{Syntactic/semantic drifting during iterative refinement.} Recent methods based on iterative refinement \citep{wang2023selfconsistencyimproveschainthought,yao2023reactsynergizingreasoningacting}, still struggle with generative control across iterations, where new inconsistencies are often introduced across new iterative cycles.
\medskip
\paragraph{(Challenge 2)}\textit{Lack of a mechanism for monotonic generalization.} Most methods focus on delivering an \textit{ab initio} mapping/solution on a query-by-query basis, without a persistent generalization which can be reused for downstream queries within a database.
\medskip
\paragraph{(Challenge 3)}\textit{Semantic mapping and large context windows.} At the conceptual center of text-to-SQL is the challenge of addressing the \textit{semantic mapping} of the natural language query referents to heterogeneous schemas, integrating multiple tables, interpreting attribute/value domain expressions over potentially large databases. A non-parsimonious use of large context windows increases the probability for performance degradation (hallucinations) and are ultimately unachievable for large databases.

\medskip

\paragraph{The Mechanism: Reflective self-correcting reasoning over localized feedback.}
\medskip

\begin{figure*}[t]
  \centering
  \includegraphics[width=\linewidth]{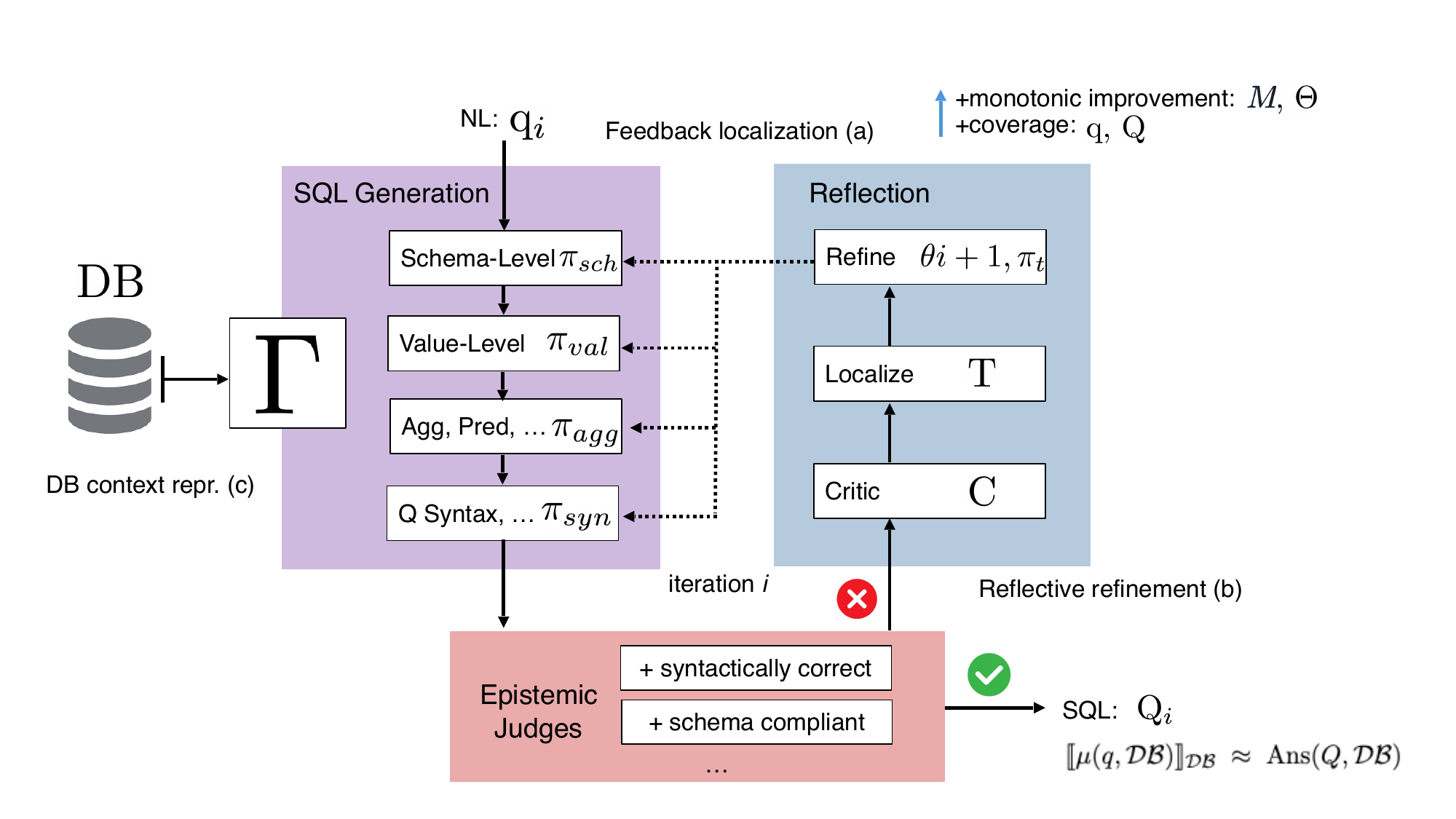}
  \caption{High-level overview of the reflective text-to-SQL approach, showing staged generation, critic feedback, semantic checking, and iterative refinement.}
  \label{fig:pipeline}
\end{figure*}

In order to address the challenges for LLM-guided text-to-SQL generation, we propose the close integration of three types of intervention:

\noindent \textbf{Feedback localization.} Addressing \textit{Challenge 1}, we decompose the text-to-SQL process into a sequence of well-defined query generation stages (Figure \ref{fig:pipeline} (a)), which can be intervened in a compartmentalized manner (rather than regenerating queries monolithically). Errors are identified, localized and diagnosed by a \textit{critic-analyzer} mechanism. 

\noindent \textbf{Reflective refinement.} Addressing \textit{Challenge 2}, we propose a reflective refinement mechanism which delivers a persistent modification of each query generation stage using the granular \textit{diagnostic context} generated by the critic-analyzer. As an ensemble mechanism, the critic--refiner loop localizes violations, reflects on which constraint failed, and selectively re-invokes only the responsible stage, enabling targeted correction without breaking previously validated semantics, allowing for a reflective monotonic improvement of the query generation process. This allows for a mechanism which abstracts over the larger semantic parsing problem.

\noindent \textbf{DB context proxy.} Complementarily, addressing \textit{Challenge 3}, we provide a database representation mechanism which disentangles database schema and value/content, in order to minimize context window use (Figure \ref{fig:pipeline} (c)), and allow for a controlled interpretation of schema-level and value-level contexts.

The mechanism defined by these components (Figure \ref{fig:pipeline}), allows for a \textit{self-correcting reflective loop} capable of abstracting over errors and delivering \textit{monotonic coverage improvements} over the query synthesis/generation process. The proposed method does not rely on supervised gold-standards for training, using proxy evaluative functions (targeting \textit{syntactic correctness}, \textit{semantic preservation}, and \textit{modular and localized error analysis}) as a feedback proxy for self-correction. Experiments on state-of-the-art relational benchmarks demonstrate consistent gains, robust convergence, and fine-grained error attribution across model families.

The contributions of this work are:
\begin{itemize}
    \item We introduce the mechanism of \textbf{reflective refinement} which can abstract over errors and refine the text-to-SQL generation process (instead of the local query).
    \item We describe and formalize the proposed approach.
    \item We provide an extensive empirical analysis of the approach using state-of-the-art benchmarks.
\end{itemize}

The remainder of the paper is structured as follows. 
Section~\ref{sec:prelim} formalizes text-to-SQL generation as constraint satisfaction and introduces the notions of monotonic refinement and semantic coverage. 
Sections~\ref{sec:refl} and~\ref{sec:db_context} present the proposed controlled multi-stage reasoning framework, including database schema interpretation, predicate grounding, query construction, and reflective refinement. 
Section~\ref{sec:experiments} describes the experimental setup and benchmarks, while Section~\ref{sec:analysis} analyzes convergence behavior, feedback granularity, and component contributions. 
Finally, Section~\ref{sec:conclusion} concludes with a discussion of limitations and future directions. All supplementary material can be found in the Appendix.

To support reproducibility, we release the full codebase, prompts, experimental configurations, and evaluation scripts used in this work, together with instructions for reproducing all reported results.\footnote{Link omitted for anonymity.}

\section{Preliminaries}
\label{sec:prelim}

\noindent\textbf{Problem definition.} Let $\mathcal{DB}$ be a relational database. Given a natural-language query $q$, the text-to-SQL task is to produce an executable SQL query $Q$ over $\mathcal{DB}$ whose execution semantics matches the intent expressed by $q$.
We model this as a mapping
\begin{equation}
\mu : \mathcal{Q} \times \mathcal{DB} \;\rightarrow\; \mathcal{L}_{\textsc{SQL}}(\mathcal{DB}),
\qquad
(q,\mathcal{DB}) \mapsto Q,
\end{equation}
where $\mathcal{Q}$ is the space of natural-language queries and $\mathcal{L}_{\textsc{SQL}}(\mathcal{DB})$ is the set of executable SQL queries over $\mathcal{DB}$.
Correctness is semantic (denotational): with execution semantics $\llbracket \cdot \rrbracket_{\mathcal{DB}}$,
\begin{equation}
\llbracket \mu(q,\mathcal{DB}) \rrbracket_{\mathcal{DB}}
\;\approx\;
\mathrm{Ans}(Q,\mathcal{DB}),
\end{equation}
where $\mathrm{Ans}(Q,\mathcal{DB})$ denotes the intended answer and $\approx$ abstracts away from surface-form differences.

\medskip
\noindent\textbf{Compositional generation.} Rather than inducing $\mu$ as a monolithic function synthesizer, we represent it as a composition of \emph{typed generative functions}
$\{\pi_t\}_{t=1}^{T}$. The overall mapping is: 
\begin{equation}
\mu \;\triangleq\; \pi_T \circ \pi_{T-1} \circ \cdots \circ \pi_1.
\label{eq:mu_comp}
\end{equation}

\medskip
\noindent\textbf{Localized generative functions.}
We assume the overall mapping $\mu$ factorizes into a composition of typed generative functions $\pi^{(t)}$, each targeting a disjoint abstraction layer of the text-to-SQL problem: $\pi^{(\textsc{sch})}$ for schema-level mappings (referent tables, attributes, join structure), $\pi^{(\textsc{val})}$ for value/instance-level grounding (attribute domains and canonical values), $\pi^{(\textsc{agg})}$ for aggregation and projection constraints, $\pi^{(\textsc{pred})}$ for predicate constraints (filters, literals, operators), and $\pi^{(\textsc{sql})}$ for the final declarative realization $q \mapsto Q$ under the syntactic constraints of the target language $\mathcal{L}$ (SQL).

\section{Reflective refinement}
\label{sec:refl}

\paragraph{Outline.}

Figure~\ref{fig:pipeline} illustrates the overall architecture. Given a natural language question $q$ and a database $\mathcal{DB}$, the system proceeds through an initial set of query generation functions, followed by a verification step. It then uses evaluative proxies to determine the syntactic and semantic correctness of the query. If violations are detected, a reflective loop diagnoses and localized the errors, through a process of reflection and abstraction, which guides a process of refinement contained to specific query generation functions, maximizing the preservation of previously validated constraints. The process terminates when the generated query satisfies both structural and semantic checks, or when a fixed refinement budget is reached. A full procedural specification of the refinement loop is provided in Appendix~\ref{app:algorithm}.





\paragraph{Reflective refinement.}
Fix a query set $\{(q_i,\mathcal{DB})\}_i$ and a current set of constraints $\Theta=\{\theta_t\}_{t\in\mathcal{T}}$ of the generators $\pi^{t}_{\theta}$ (hence $\mu_\Theta$). Please note that ${\theta_t}$ is a prompt component associated with the typed generative function $\pi^{t}$. 

\noindent Each refinement cycle proceeds as follows:
\begin{enumerate}[leftmargin=*,nosep]
\item \textbf{Generate:} for each $i$, produce $Q_i \leftarrow \mu_\Theta(q_i,\mathcal{DB}_i)$.
\item \textbf{Evaluate:} obtain natural-language diagnostics
$
r_i \leftarrow \mathrm{eval}(Q_i,\mathcal{DB}_i,q_i),
$
where $\mathrm{eval}$ includes a syntactic/execution component and a semantic (LLM) intent-preservation component, i.e. $[[\mu(q,\mathcal{DB})]]_{\mathcal{DB}}
\;\approx\;\mathrm{Ans}(Q,\mathcal{DB})$.

\item \textbf{Critique:} identify the query inconsistencies:
$
\{(t,\,c,\,Q_i')\} \leftarrow f_{critic}(r_i, \Gamma_{\mathcal{DB}},Q_i)
$
\item \textbf{Localize:} identify the target component type:
$
\pi_t \leftarrow \ell(r_i)\in\mathcal{T}.
$
\item \textbf{Reflect-refine:} update only the implicated parameters in $\pi_t$:
$
\theta_{t,i+1} \leftarrow \mathrm{Reflect}(\theta_{t},i; r_i),
$
\item \textbf{Maximize backwards preservation:} while leaving all $\theta_{t},i$ semantically preserved.
\end{enumerate}

\medskip
The essential distinction from standard refinement is that the edits are applied to the \emph{mechanism} parameters $\theta_t$ (e.g., prompts of $\pi^{(t)}$), not to the specific query instance $Q_i$.

\medskip
\noindent\textbf{Reflect.} After a query hypothesis $Q_i$ construction, the system evaluates the candidate query using a \emph{critic-refiner} loop. The \textbf{critic} is an in-context LLM-based interpretation function $f_{critic}$ which performs a systematic criteria-based ($\mathcal{C}$) inspection of the query, against the $\mathcal{DB}$ context proxy $\Gamma$ (see Appendix \ref{app:prompt_semantic_plan}). The set of criteria $\mathcal{C}$ are defined by a set of critical evaluative functions according to each type $\mathcal{T}$: (i) schema-level violations, i.e. invalid table and attribute referents, joins); (ii) value-level violations, i.e. identifier and formatting inconsistencies for content-level referents; as well as (iii) aggregation and predicate semantics and (iv) query syntax violations. The critic returns a localized violation set
$\{(t,\,c,\,Q_i')\}, t\in\mathcal{T},\; c\in\mathcal{C}_t,\; Q_i' \subseteq Q_i$, where $t$ is the violated abstraction type, $c$ the specific criterion, and $Q'$ the implicated query fragment (e.g., span/clause). The corresponding critic implementation is covered in Appendix \ref{app:critic-prompt}.

\medskip
\noindent\textbf{Abstract \& Refine.} This selective restart mechanism contrasts with generic retry strategies: it uses the localized diagnostic input of the previous step, to abstract over the set of parameters $\theta_i$ in $\pi_t$, which have a coverage of a query set ${q_i,Q_i}$, updating them ($\theta_{i+1}$) within a query generation component ($\pi_t$) while maximizing preservation of correct reasoning steps. The corresponding refine function implementation is covered in Appendix \ref{app:algorithm}.

\medskip
\noindent\textbf{Unsupervised feedback via epistemic judges.} Because no gold SQL is assumed, progress is defined only through $\mathrm{eval}$ as a proxy function of syntactic correctness and semantic preservation. This work adapts the recently introduced concept of formal epistemic judges \citep{} to the domain of query synthesis. Initially introduced in the area of mathematical auto-formalization, formal epistemic judges aims at providing an ensemble of evaluative functions which can serve as a granular evaluative proxy for complex tasks. Epistemic judges can be LLM-based (L) on interpreter-based (I). In the context of this work, the implementation of the evaluative functions for \textit{syntactic correctness} (I), \textit{semantic preservation} (between ${q_i, Q_i}$) (L).

\section{DB Context Interpretation}
\label{sec:db_context}

Relational databases $\mathcal{DB}$ are typically too large and heterogeneous to be provided to an LLM ``as is'': the full schema can be long, and the instance-level content is effectively unbounded. For controlled text-to-SQL generation we therefore replace $\mathcal{DB}$ by a compact, reusable \emph{context proxy} $\Gamma$ that (i) preserves the schema constraints that govern what is \emph{legal} to generate, and (ii) exposes only a carefully budgeted sample of instance-level evidence needed for grounding the value domains of the attributes. Concretely, we construct a cached context $\Gamma$ as an offline preprocessing step; $\Gamma$ is the interface through which downstream generators $\pi_t$ access $\mathcal{DB}$.

\medskip
\noindent\textbf{Context proxy.}
Given a database $\mathcal{DB}$, we define a context construction mapping
\begin{equation}
\Gamma \;=\; \Gamma(\mathcal{DB}),
\end{equation}
where $\Gamma$ bundles schema-level and value-level interpretive/sampling functions. Intuitively, $\Gamma$ is a compact representation that supports controlled access to (a) schema structure and (b) limited value evidence, without requiring end-to-end encoding of the entire database.

\medskip
\noindent\textbf{Schema-level interface.}
$\Gamma$ provides functions that normalize and summarize schema elements and constrain feasible joins:
\begin{equation}
\Gamma_{\textsc{sch}} \;=\; \big(f_{\Delta},\;f_{\Sigma},\;f_{\mathcal{J}}\big),
\end{equation}
where $f_{\Delta}$ produces normalized table/column descriptors (canonical names, types, key likelihood), $f_{\Sigma}$ produces concise natural-language table summaries to support scalable table selection, and $f_{\mathcal{J}}$ produces join candidates derived from declared foreign keys (optionally corroborated by samples) to restrict feasible join paths.

\medskip
\noindent\textbf{Value-level interface (budgeted evidence).}
When samples are available, $\Gamma$ additionally provides a cardinality-aware value grounding interface:
\begin{equation}
\Gamma_{\textsc{val}} \;=\; f_{\mathcal{V}},
\end{equation}
where $f_{\mathcal{V}}$ exposes representative values for low-cardinality attributes (e.g., status/category fields) while suppressing enumeration for high-cardinality attributes (e.g., names) to keep context small. This yields a controlled lexicon for literal grounding without expanding the prompt with large value lists.

\medskip
\noindent\textbf{Usage.}
Downstream stages consume $\Gamma$ as the sole database-facing context: (a) table/attribute selection is guided by normalized descriptors and summaries, (b) joins are restricted to candidates licensed by $f_{\mathcal{J}}$, and (c) literal/constraint grounding is improved by $f_{\mathcal{V}}$ under a fixed budget. In this sense, $\Gamma(\mathcal{DB})$ serves as a practical proxy for $\mathcal{DB}$, providing enough evidence for accurate generation while enforcing controllability and reuse across a query set.

\section{Experiments}
\label{sec:experiments}

We evaluate \method{} on two multi-domain text-to-SQL benchmarks, \datasetSpider{} and \datasetBird{}, to test (i) end-to-end execution accuracy, (ii) convergence under iterative refinement, and (iii) the contribution of feedback and semantic coverage checking.

\subsection{Setup}
\label{subsec:exp_setup}

\paragraph{Benchmarks.}
We use the official development splits of \datasetSpider{} and \datasetBird{}. The \datasetSpider{} dev set contains 1,034 questions spanning 20 databases drawn from the benchmark’s pool of 200 cross-domain schemas, emphasizing compositional generalization over heterogeneous structures. \datasetBird{} consists of 12,751 total pairs across 95 databases, and adds more challenging SQL patterns and evaluates both query validity and efficiency, providing a complementary stress test for complex relational reasoning.

\paragraph{Baselines.}
We compare against strong GPT-4 based prompting systems: DIN-SQL \citep{pourreza2023dinsql}, DAIL-SQL \citep{gao2023dailsql}, MAC-SQL \citep{wang2025macsql}, and MCS-SQL \citep{lee2024mcssql}, as well as a GPT-4 zero-shot baseline \citep{achiam2023gpt4}.

\paragraph{Models and decoding.}
We instantiate \method{} with a representative set of foundation models spanning both proprietary and open-weight regimes. GPT-4 and GPT-5 class models serve as strong upper-bound references, capturing the performance of state-of-the-art general-purpose LLMs under controlled prompting, and to compare ours to previous SOTA approaches. In parallel, we evaluate competitive open-weight models in the 30B parameter range (e.g., Qwen3-30B and DeepSeek-R1-32B), which offer a favorable trade-off between capacity and accessibility and are commonly used in recent text-to-SQL systems.

To test whether the proposed mechanism relies on scale or persists under constrained capacity, we additionally run targeted ablations using a substantially smaller open-weight model (LLaMA~3.1~8B). This setting allows us to verify that the benefits of controlled decomposition, semantic planning, and selective refinement extend beyond frontier-scale models and remain effective even when base model reasoning is limited.

Unless noted otherwise, all stages use deterministic decoding (temperature $\approx 0$) with fixed token budgets per stage to reduce variance and ensure comparability across models.

\paragraph{Refinement budget.}
We allow up to $T$ refinement steps per example with early stopping when structural checks and semantic coverage checks pass. The refiner selectively restarts only the stage implicated by the critic/analyzer signals (schema interpretation, predicate grounding, or query construction).

\paragraph{Metrics.}
We report execution accuracy (EX) on both benchmarks and Valid Efficiency Score (VES) on \datasetBird{}.

\subsection{Main Results}
\label{subsec:main_results}



\begin{table}[t!]
  \centering
  \small
  \begin{tabularx}{\linewidth}{
    >{\raggedright\arraybackslash}X
    S[table-format=2.1]
    S[table-format=2.1]
    S[table-format=2.1]
  }
    \toprule
    \textbf{Method} &
    \multicolumn{2}{c}{\textbf{BIRD}} &
    \multicolumn{1}{c}{\textbf{Spider}} \\
    \cmidrule(lr){2-3}\cmidrule(lr){4-4}
    & \multicolumn{1}{c}{\textbf{VES}} & \multicolumn{1}{c}{\textbf{EX}} & \multicolumn{1}{c}{\textbf{EX}} \\
    \midrule
    GPT-4 (zero-shot)                & 49.8 & 46.4 & 74.6 \\
    DIN-SQL                          & 58.8 & 50.7 & 82.8 \\
    DAIL-SQL                         & 56.1 & 54.8 & 84.4 \\
    MAC-SQL                          & 58.8 & 57.7 & 86.8 \\
    MCS-SQL                          & 64.8 & 63.4 & 89.5 \\
    \midrule
    \method{} (base)                 & 58.5 & 54.6 & 81.7 \\
    \method{} (full)                 & 66.5 & 65.2 & 93.8 \\
    \addlinespace[0.25em]
    \quad--- GPT-5                    & {\bfseries 68.9} & {\bfseries 67.8} & {\bfseries 95.4} \\
    \quad--- Qwen3-30B                & 61.2 & 59.9 & 92.1 \\
    \quad--- DeepSeek-R1-32B          & 63.7 & 62.1 & 94.3 \\
    \bottomrule
  \end{tabularx}
  \caption{%
  \textbf{The full \method{} consistently outperforms prior prompting and agentic baselines, demonstrating the benefit of reflective refinement beyond staged decomposition alone.}
    Baseline and SOTA comparison on \datasetSpider{} and \datasetBird{}, reporting valid efficiency score (VES) and execution accuracy (EX).
    All methods use GPT-4 as the foundation model unless otherwise specified for \method{} variants.
  }
  \label{tab:sota}
\end{table}

\textbf{Across all benchmarks, \method{} outperforms all prompting baselines and prior agentic systems.} The full method including feedback-refinement iterations consistently outperforms the staged baseline, indicating that reflective refinement and targeted restarts add substantial value beyond decomposition alone. Gains persist across model families, suggesting the method is not tied to a specific foundation model.

\section{Analysis}
\label{sec:analysis}

\subsection{Convergence under Refinement}
\label{subsec:analysis_convergence}

\begin{figure*}[t]
 \centering
   \includegraphics[width=\textwidth]{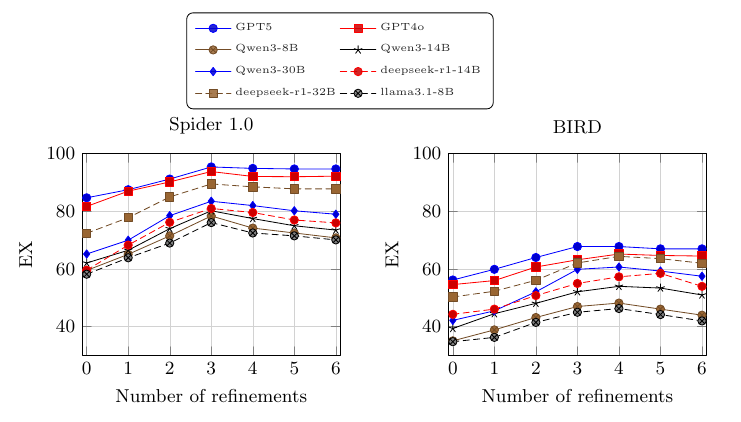}
   \caption{%
\textbf{Early refinement yields the majority of gains across all models, while excessive iterations lead to diminishing returns or over-refinement, motivating selective and bounded refinement.}
   EX is reported at each refinement step $t$ for a range of base models on \datasetSpider{} (dev) and \datasetBird{} (dev).
 }
 \label{fig:iteration_ex_models_two}
\end{figure*}

\textbf{Execution Accuracy (EX) consistently increases with the increase of the refinement budget across all foundation models and datasets.} 
Figure~\ref{fig:iteration_ex_models_two} plots EX as a function of refinement iterations. An average of $\sim12$ percent in EX occur within the iterations $t=0$ to $t=3$ on the \datasetSpider{} (from $\sim85$ to $\sim95$ EX for GPT5, and from $\sim82$ to $\sim94$ EX for GPT4o) after which performance saturates and can degrade for weaker models (see Llama3.1-8B). Medium-sized models like Qwen3-30B and \texttt{deepseek-r1-32B} show even larger relative gains (e.g., \texttt{deepseek-r1-32B} improves from 72.4 to 89.5 EX by $t=3$), indicating that the agentic refinement loop is particularly beneficial when the single-shot performance is still far from saturation. Smaller models such as Qwen3-8B and \texttt{llama3.1-8B} also benefit substantially in the first three iterations (e.g., Qwen3-8B: 59.5 to 78.4 EX by $t=3$), but their curves exhibit more volatility after $t=3$ than seen in larger models. 

Beyond $t=3$, additional iterations on \datasetSpider{} yield diminishing returns and can even hurt performance. For the strongest models (GPT5, GPT4o, \texttt{deepseek-r1-32B}), EX remains consistent or slightly decreasing after $t=3$, suggesting that the model has already converged and later edits mostly produce benign perturbations. In contrast, weaker models show clear signs of over-refinement: Qwen3-8B and \texttt{llama3.1-8B} drop by 5--8 EX points between $t=3$ and $t=6$. Qualitatively, we find cases where an early iteration introduces a hallucinated table, join, or predicate; subsequent iterations then reflect on and refine this mistaken hypothesis rather than reverting it, causing the system to ``lock in'' to an incorrect execution plan.

The behaviour on \datasetBird{} (right) mirrors these trends but with lower absolute scores and a slightly delayed drop-off. All models start significantly below their \datasetSpider{} counterparts at $t=0$ (e.g., GPT5 at 56.2 EX and GPT4o at 54.6 EX), reflecting the higher schema complexity and SQL difficulty of \datasetBird{}. However, the first three to four refinement steps again provide substantial improvements: GPT5 reaches 67.8 EX by $t=3$ and remains stable thereafter, while GPT4o climbs to 65.2 EX by $t=4$. Larger open-source models (Qwen3-30B, \texttt{deepseek-r1-32B}) follow a similar trajectory, with peaks around $t=3$--$4$ and mild degradation afterwards. Smaller models (Qwen3-8B, \texttt{llama3.1-8B}) gain 10--15 EX points up to $t=3$, but their performance starts to noticeably degrade for $t \ge 5$, again consistent with over-refinement around brittle or hallucinated intermediate beliefs.

Overall, these curves suggest that (i) a small refinement budget ($t \approx 3$ on \datasetSpider{} and $t \approx 3$--$4$ on \datasetBird{}) captures most of the available gains, (ii) increasing the iteration budget beyond this range offers little benefit for strong models and can harm weaker ones, and (iii) the agentic loop systematically narrows, but does not erase, the gap between small and large models. In practice, this points to using a moderate, model- and benchmark-specific iteration cap rather than running the refinement loop to exhaustion.

\subsection{Feedback Granularity}
\label{subsec:analysis_feedback}

\begin{table}[t]
  \centering
  \small
  \begin{tabularx}{\linewidth}{l S[table-format=2.1] S[table-format=2.1] S[table-format=1.1]}
    \toprule
    \textbf{Feedback} & \textbf{BIRD} & \textbf{Spider} & \textbf{Avg. Iters} \\
    \textbf{Style} & \textbf{(EX)} & \textbf{(EX)} & \textbf{(Spider)} \\
    \midrule
    Coarse          & 61.0 & 90.7 & 2.3 \\
    Granular        & \textbf{65.2} & \textbf{93.8} & \textbf{3.1} \\
    Epistemic-only  & 64.7 & 92.5 & 2.6 \\
    \bottomrule
  \end{tabularx}
  \caption{%
    \textbf{Localized, categorical feedback enables more precise corrections and higher execution accuracy than coarse or epistemic-only feedback, while requiring slightly more refinement iterations.}
    Comparison between coarse and granular categorical feedback for a fixed base model (GPT4o) on \datasetSpider{} and \datasetBird{}.
  }
  \label{tab:feedback_styles}
\end{table}

\textbf{Highly localized, categorical feedback consistently improves execution accuracy, at the cost of a small increase in refinement iterations.} Table~\ref{tab:feedback_styles} compares three feedback styles for a fixed base model (GPT4o) on \datasetBird{} and \datasetSpider{}. Coarse feedback, which only marks an entire candidate as correct or incorrect with minimal localization of the error, already yields strong performance (61.0 and 90.7 EX respectively) with 2.3 refinement steps on average. Granular feedback, consisting of category-specific and span-level signals (e.g. pointing to incorrect joins, predicates, or missing constraints), leads to a consistent improvement on both benchmarks, over the coarse feedback, at the cost of slightly more refinement steps (3.1 iterations). This is expected: more informative feedback allows the model to make more targeted corrections per step, but also encourages continuous refinement until a higher-quality solution is reached.

Epistemic-only feedback, which focuses on correcting the model's underlying assumptions and high-level generalizations about the schema interpretation and question (e.g. mismatched entity types or misinterpreted relations) without providing fully granular structural hints, sits between the two extremes, achieving 64.7 and 92.5 EX respectively, at only 2.6 refinement iterations on average.

Overall, granular feedback enables more reliable correction by pinpointing which semantic constraints must be revised, rather than triggering global regeneration.

\subsection{Component Contributions}
\label{subsec:analysis_ablation}


\begin{table}[t]
  \centering
  \small
  \begin{tabularx}{\linewidth}{l S[table-format=2.1] S[table-format=2.1] S[table-format=2.1] S[table-format=2.1]}
    \toprule
    \textbf{Configuration} & \textbf{VES} & \textbf{EX} & \textbf{$\Delta$VES} & \textbf{$\Delta$EX} \\
    \midrule
    Full method          & \textbf{66.5}  & \textbf{65.2}  &  &  \\
    No critic (syntax-only)& 62.4 & 61.6 & -4.1 & -3.6 \\
    No semantic checker    & 63.1 & 58.2 & -3.4 & -7.0 \\
    Single-shot (no stages)& 58.5 & 54.6 & -8.0 & -10.6 \\
    \bottomrule
  \end{tabularx}
    \caption{%
    \textbf{Accuracy drops when removing staged decomposition, the critic--refiner loop, or the semantic checker}, showing that robust gains arise from their interaction rather than any single component in isolation. Component ablations on \datasetBird{}.
    Each row removes one or more components from the full pipeline; $\Delta$ columns show performance changes relative to the full system.
  }
  \label{tab:ablations}
\end{table}
\textbf{Performance gains arise from the combination of staged decomposition, localized refinement, and intent-level semantic verification, rather than from any single component alone.}
To isolate where gains come from, we ablate major components (Table~\ref{tab:ablations}). Removing the critic--refiner loop leads to a noticeable drop in both VES and EX, indicating that execution-grounded feedback and diagnostics are critical for correcting generated SQL queries. Disabling the semantic checker causes an even larger degradation (-7.0 EX compared to full method), highlighting the importance of intent-level verification beyond syntax and execution. Finally, collapsing the system into a single-shot generator without staged constraints yields the largest performance loss, confirming that decomposition itself is a necessary (but insufficient) condition for robust refinement. 

Taken together, this shows that reliable improvement emerges from the interaction between constraint-emitting stages and targeted reflective refinement, rather than from decomposition or verification in isolation.

\subsection{Qualitative Cases and Benchmark Limitations}
\label{subsec:analysis_qualitative}

\textbf{Many apparent text-to-SQL ``errors'' arise from evaluation underspecification rather than genuine semantic failure, underscoring the need for intent-level verification.}
Exact-match evaluation conflates true semantic failures with (i) alternative-but-valid SQL forms and (ii) questionable gold annotations in the \datasetSpider{}. Table~\ref{tab:qualitative} illustrates three recurring patterns.
First, \textbf{schema/join ambiguity}: multiple join paths can yield equivalent denotations, but benchmarks often provide a single canonical query.
Second, \textbf{boundary and literal errors}: true failures are frequently localized to operator choice or literal grounding (e.g., $>$ vs.\ $\ge$), which is precisely the type of error our staged design can attribute to predicate grounding.
Third, \textbf{over-/under-specification}: model outputs may include plausible constraints absent from gold SQL (or omit constraints that gold SQL implicitly assumes), exposing a mismatch between benchmark minimalism and practical intent interpretation.
These observations motivate semantic coverage checks that evaluate whether required constraints are present, independent of surface form.

\section{Related Work}

\noindent\textbf{Schema linking and grounding.} Aligning mentions to schema/value evidence remains a core bottleneck. LinkAlign scales schema linking to large multi-database settings \citep{wang-etal-2025-linkalign}, while contextual schema-link graphs improve ICL demo selection and grounding \citep{lee-etal-2025-dcg}. Content-aware prompting (e.g., value-grounded question rewriting) further reduces ambiguity before generation \citep{mao-etal-2024-enhancing}.

\noindent\textbf{Decomposition, planning, and structured in-context learning.} Many systems improve reliability via stepwise decomposition plus self-correction. DIN-SQL operationalizes decomposed ICL with iterative correction \citep{DBLP:conf/nips/PourrezaR23}; PTD-SQL partitions by query type for targeted ``drilling'' \citep{luo-etal-2024-ptd}; ROUTE uses robust multitask tuning/collaboration to stabilize cross-subtask generalization \citep{qin-etal-2025-route}.

\noindent\textbf{Interaction and multi-turn text-to-SQL.} Conversational text-to-SQL requires context carryover and incremental edits; CoE-SQL models this as a chain of SQL editions over prior queries \citep{zhang-etal-2024-coe}.

\noindent\textbf{Execution-guided verification and agentic refinement loops.} Execution feedback is increasingly central for refinement and supervision. DART-SQL repairs candidates with execution-guided refinement \citep{mao-etal-2024-enhancing}, while ExeSQL bootstraps self-taught data across SQL dialects using execution validation \citep{zhang-etal-2025-exesql}. Multi-candidate reasoning with preference-optimized selection further improves final execution accuracy \citep{pourreza-etal-2025-chase}.

\noindent\textbf{Data synthesis and fine-tuning strategies for open models.} Closing the open--closed gap relies on synthetic supervision and targeted tuning. DTS-SQL uses staged decomposition for small LLMs \citep{pourreza-rafiei-2024-dts}, and weak/strong LLM synthesis pipelines provide scalable training data for instruction tuning \citep{yang-etal-2024-synthesizing}.

\noindent\textbf{Positioning of this work.}
Across these threads, prior systems primarily improve text-to-SQL by strengthening grounding (schema/value linking), decomposing generation into stages, and then applying feedback by \emph{repairing or regenerating the current SQL instance}, or by introducing additional supervision via synthesis/tuning. In contrast, our core contribution is to treat refinement as an update to the \emph{stage-level generation mechanism} (prompt/parameter updates of typed generators) driven by localized diagnostics and abstraction and with the support of epistemic proxies (without assuming gold SQL or additional training), so improvements \emph{transfer across a query set} within a database rather than remaining brittle per-query edits. 

\noindent

\section{Conclusion}
\label{sec:conclusion}

This work reframes iterative improvement in text-to-SQL as \emph{reflective} refinement mechanism, rather than instance-level SQL rewriting. Starting from the three core failure modes of current LLM-guided pipelines—drift under refinement, lack of transferable generalization, and poor scalability of schema/value grounding—we introduced a controlled, staged generation process coupled to a Reflect--Refine Loop that localizes violations. The central idea of \emph{reflective refinement} is to update the parameters of the stage generators (e.g., prompt constraints) using localized diagnostic context, while maximizing backward preservation of already-validated constraints, enabling monotonic coverage improvements over a query set within a database. Complementarily, we proposed a compact, reusable database context proxy that disentangles schema structure from sampled value/domain evidence, reducing context-window dependence and improving controlled grounding.

Within the scope of the presented design, the resulting system yields a self-correcting loop that is both more controllable and more reusable than monolithic retry strategies: refinement targets the smallest responsible abstraction layer, and progress is assessed via unsupervised epistemic judges combining interpreter-based syntactic/execution checks with intent-level semantic coverage verification. This combined mechanism provides a practical pathway to robust text-to-SQL behavior without assuming gold SQL supervision, while remaining compatible with diverse foundation models. Beyond text-to-SQL, the mechanism suggests a general template for structured prediction in which iterative improvement is driven by localized diagnostics and constraint-preserving updates to the underlying generative operators.


\section*{Limitations}

The proposed approach is bounded by a fixed refinement budget and returns the final SQL candidate even if issues persist once this budget is exhausted, which may leave some difficult cases unresolved. Its effectiveness depends on the quality of the Critic and Analyzer components: inaccurate or coarse diagnostics misattribute failures to the wrong abstraction layer, leading to inefficient refinement or missed corrections. 
Moreover, the current semantic checker enforces a predefined set of correctness constraints (e.g., schema consistency, join feasibility, aggregation semantics) and may fail to detect subtler semantic mismatches that remain executable but deviate from the user’s intent. While the framework is intentionally designed around semantic faithfulness and controllability, it does not explicitly reason about query efficiency or execution cost. Incorporating efficiency-oriented epistemic judges—e.g., based on query plans or execution statistics—remains an important direction for future work. 
Finally, although demonstrated here for NL-to-SQL, the framework is general. Applying it to other structured prediction tasks (e.g., text-to-SPARQL) would require task-specific instantiations of generators and evaluators, rather than changes to the core reflective refinement mechanism.

\section*{Acknowledgments}

This work was supported by Merck KGaA as part of an academic–industrial collaboration. We thank our colleagues at the Idiap Research Institute as well as colleagues at Merck KGaA for valuable discussions and suggestions that helped shape this work.

\bibliography{custom}

\appendix
\clearpage
\onecolumn

\clearpage
\section{Error Analysis}
\label{app:error_analysis}

This appendix provides a qualitative error analysis intended to contextualize the quantitative results reported in the main paper. The examples in Table~\ref{tab:qualitative} illustrate representative model behaviors across different query complexities, including true semantic errors, alternative-valid predictions, and cases where the benchmark gold annotation is incorrect or underspecified.
We include both correct and incorrect predictions to highlight how our system’s stage-wise design affects failure modes, particularly in the presence of aggregation, joins, and schema ambiguity.

Each row reports the natural-language question, the gold SQL annotation, the model’s predicted SQL, a coarse complexity category, and the final outcome label.
Outcome labels follow our evaluation protocol: \emph{Alternative-valid} denotes queries that differ syntactically from the gold SQL but are semantically equivalent; \emph{Gold wrong} indicates errors in the reference annotation rather than the model output. This analysis is not exhaustive, but serves to illustrate common error patterns and limitations of both the model and the benchmark.

\begin{table*}[h]
  \centering
    \begin{tabularx}{\linewidth}{
    >{\raggedright\arraybackslash}X
    >{\raggedright\arraybackslash}X
    >{\raggedright\arraybackslash}X
    l
    l
  }
    \toprule
    \textbf{Question} & \textbf{Gold SQL} & \textbf{Predicted SQL} & \textbf{Complexity} & \textbf{Outcome} \\
    \midrule
    List the names of students enrolled in courses taught by Professor Smith
    &
    \texttt{SELECT s.name FROM student s JOIN enrollment e JOIN course c JOIN professor p WHERE p.name = 'Smith'}
    &
    \texttt{SELECT s.name FROM student s JOIN enrollment e JOIN teaching t JOIN professor p WHERE p.name = 'Smith'}
    &
    Medium
    &
    Alternative-valid
    \\
    \addlinespace
    Find the number of employees hired after 2019
    &
    \texttt{SELECT COUNT(*) FROM employee WHERE hire\_year > 2019}
    &
    \texttt{SELECT COUNT(*) FROM employee WHERE hire\_year >= 2019}
    &
    Easy
    &
    True error
    \\
    \addlinespace
    What are the titles of movies directed by Christopher Nolan?
    &
    \texttt{SELECT m.title FROM movie m JOIN director d WHERE d.name = 'Christopher Nolan'}
    &
    \texttt{SELECT m.title FROM movie m JOIN director d WHERE d.full\_name = 'Christopher Nolan'}
    &
    Easy
    &
    Alternative-valid
    \\
    \addlinespace
    List departments with more than five employees
    &
    \texttt{SELECT d.name FROM dept d JOIN emp e GROUP BY d.id HAVING COUNT(*) > 5}
    &
    \texttt{SELECT d.name FROM dept d JOIN emp e GROUP BY d.id, d.name HAVING COUNT(e.id) > 5}
    &
    Medium
    &
    Alternative-valid
    \\
    \addlinespace
    Find the average salary of employees in the sales department
    &
    \texttt{SELECT AVG(e.salary) FROM emp e}
    &
    \texttt{SELECT AVG(e.salary) FROM emp e JOIN dept d WHERE d.name = 'Sales'}
    &
    Easy
    &
    Gold wrong
    \\
    \bottomrule
  \end{tabularx}
    \caption{%
    \textbf{Qualitative examples and benchmark limitations.}
    For each case, we show the question, gold SQL, predicted SQL, complexity category, and outcome.
  }
    \label{tab:qualitative}

\end{table*}
\FloatBarrier

\clearpage

\section{Illustrative Example: Selective Reflective Refinement}
\label{sec:appendix_example_refinement}

\begin{figure*}[t]
  \centering
  \includegraphics[width=\linewidth]{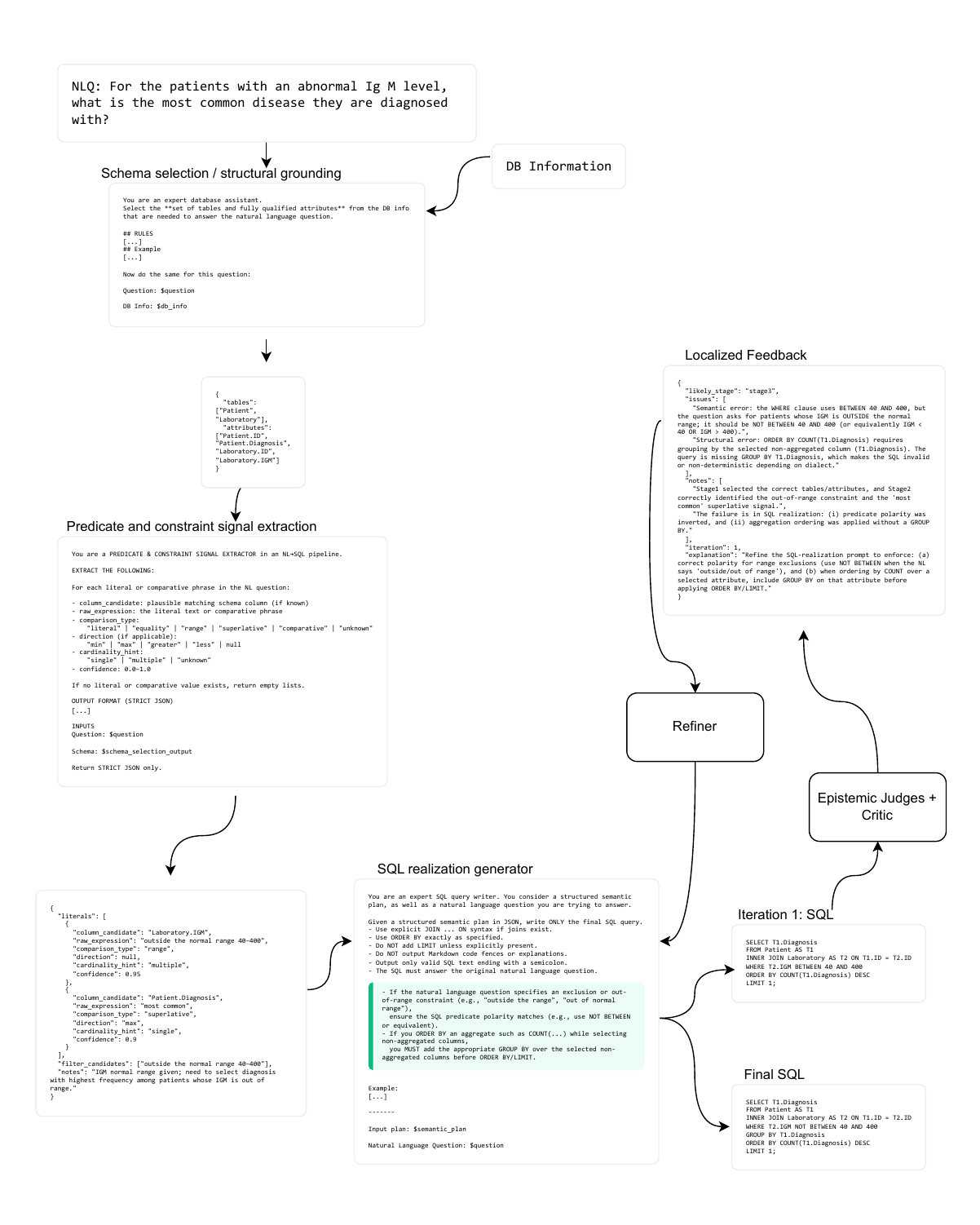}
  \caption{\textbf{Illustrative example of selective reflective refinement in the text-to-QL pipeline.}
The diagram shows an end-to-end inference trace for a query involving an out-of-range laboratory constraint and a superlative aggregation. An initial compositional pass produces an incorrect SQL realization. Epistemic judges and a critic localize the failure to the SQL realization component, triggering a targeted refinement of its generation parameters. The refined realization enforces correct predicate polarity and aggregation grouping, yielding a syntactically and semantically correct final query.}
\label{fig:appendix_reflective_example}
\end{figure*}

Figure~\ref{fig:appendix_reflective_example} visualizes a representative inference episode in the proposed controlled NL$\rightarrow$SQL framework. The pipeline first decomposes the natural-language question into schema-level grounding and predicate-level constraint signals, which are then realized as an executable SQL query. In the first iteration, the generated query fails to preserve the intended semantics due to an inverted range predicate and a missing grouping constraint under aggregation. 

Using interpreter-based checks and LLM-based semantic evaluation, the critic attributes the error to the SQL realization component. Rather than regenerating the query instance, the system performs a selective reflective update to the parameters of the realization generator, strengthening realization-level constraints while preserving previously validated schema and predicate decisions. Re-invoking only the affected component yields a corrected SQL query that satisfies both syntactic correctness and semantic intent preservation.

\section{Algorithm}
\label{sec:appendix_algorithm}

This appendix provides pseudocode for the controlled, compositional text-to-SQL inference procedure described in the main paper. The Algorithm~\ref{app:algorithm} formalizes the interaction between the typed query generation components corresponding to distinct abstraction layers (e.g., schema-level, predicate-level, aggregation-level, and SQL realization), and the selective refinement loop applied during inference.

The algorithm distinguishes between an intial compositional generation pass, instantiating the mapping $\mu_{\Theta}=\pi_{|\mathcal{T}|}\circ\cdots\circ\pi_1$ and a refinement loop in which only implicated generator components are selectively re-invoked. Refinement is guided by a Critic that produces localized violation diagnostics and attributes errors to a specific abstraction type, enabling targeted updates to the parameters of the corresponding generator while preserving previously validated constraints.

The inference process terminates early when both syntactic correctness and semantic intent preservation are satisfied, as determined by unsupervised epistemic judges, or after a fixed refinement budget is reached, in which case the best-effort executable SQL query is returned.

\begin{algorithm}[h]
\caption{Controlled compositional text$\rightarrow$SQL with selective reflective refinement}
\label{app:algorithm}
\begin{algorithmic}[1]
\Require NL question $q$, relational database $\mathcal{DB}$, max refinements $T{=}3$
\Require Typed generators $\{\pi_t\}_{t\in\mathcal{T}}$ with parameters $\Theta=\{\theta_t\}_{t\in\mathcal{T}}$
\State Initialize state $s \leftarrow (q, DB)$
\Statex

\State \Comment{Initial compositional generation: $\mu_{\Theta}=\pi_{|\mathcal{T}|}\circ\cdots\circ\pi_1$}
\State $s \leftarrow \pi_{\textsc{SCH}}(s;\theta_{\textsc{SCH}})$
\State $s \leftarrow \pi_{\textsc{VAL}}(s;\theta_{\textsc{VAL}})$
\State $s \leftarrow \pi_{\textsc{AGG}}(s;\theta_{\textsc{AGG}})$
\State $s \leftarrow \pi_{\textsc{PRED}}(s;\theta_{\textsc{PRED}})$
\State $\widehat{Q} \leftarrow \pi_{\textsc{SQL}}(s;\theta_{\textsc{SQL}})$
\Statex

\For{$i \leftarrow 1$ to $T$}
  \State \Comment{Unsupervised evaluation via epistemic judges (interpreter + LLM)}
  \State $r \leftarrow eval(\widehat{Q}, DB, q)$
  \State $\text{syn} \leftarrow Pass_{\textsc{syn}}(r)$ \Comment{parses/executes under $DB$}
  \State $\text{sem} \leftarrow Pass_{\textsc{sem}}(r)$ \Comment{intent preserved: $[[\widehat{Q}]]_{DB}\approx Ans(q, DB)$}
  \If{$\text{syn} \land \text{sem}$}
    \State \textbf{return} $\widehat{Q}$
  \EndIf
  \Statex

  \State \Comment{Critique: produce localized violation triples $(t,c,Q')$}
  \State $\Gamma_{DB} \leftarrow BuildContextProxy(DB)$
  \State $\mathcal{V} \leftarrow f_{\textsc{critic}}(r, \Gamma_{DB}, \widehat{Q})$
  \State \Comment{$\mathcal{V}=\{(t,c,Q')\}$ where $t\in\mathcal{T}$, $c\in\mathcal{C}_t$, $Q'\subseteq\widehat{Q}$}
  \Statex

  \State \Comment{Localize to the implicated abstraction type}
  \State $t^\star \leftarrow \ell(\mathcal{V})$ \Comment{$t^\star\in\mathcal{T}$}
  \Statex

  \State \Comment{Reflect-refine: update only the parameters of the implicated generator}
  \State $\theta_{t^\star} \leftarrow Reflect(\theta_{t^\star}; \mathcal{V}, r)$
  \State \Comment{Maximize backward preservation: for all $t\neq t^\star$, keep $\theta_t$ unchanged}
  \Statex

  \State \Comment{Selective restart: re-run only the implicated component (and downstream realization)}
  \State $s \leftarrow \pi_{t^\star}(s;\theta_{t^\star})$
  \State $\widehat{Q} \leftarrow \pi_{\textsc{SQL}}(s;\theta_{\textsc{SQL}})$
\EndFor

\State \Return $\widehat{Q}$ \Comment{budget exhausted: return best-effort SQL}
\end{algorithmic}
\end{algorithm}
\FloatBarrier
\section{Prompts}
\subsection{Stage 1: Table and Attribute Selection Prompt}
\label{app:prompt_schema}

\paragraph{Purpose.}
This prompt performs schema-level interpretation by selecting the minimal set of tables
and fully qualified attributes required to answer the input question.
It emits schema constraints that determine the permissible join structure
and attribute footprint for downstream reasoning stages.

\begin{tcolorbox}[title={Stage~1 — Table and Attribute Selection (SYSTEM PROMPT)},
  breakable
]
\textbf{Role.}
You are an \textbf{expert database assistant} responsible for \emph{schema-level interpretation}.
Given a natural-language question and a database schema, select the \emph{minimal} set of tables
and \emph{fully qualified attributes} required to express a correct SQL query.

\tcblower

\textbf{Objectives}
\begin{enumerate}
  \item \textcolor{blue}{Identify the minimal set of relevant tables} needed to answer the question.
  \hfill {\scriptsize [Schema Interpretation]}
  \item \textcolor{teal}{Select only the attributes required} to express joins, filters, grouping, or outputs.
  \hfill {\scriptsize [Attribute Footprint Minimization]}
  \item \textcolor{purple}{Emit explicit schema constraints} to guide downstream reasoning stages.
  \hfill {\scriptsize [Structural Constraint Inference]}
\end{enumerate}

\textbf{Critical Requirements}
\begin{itemize}
  \item \textbf{Attribute qualification.}
  All attributes \emph{must} be fully qualified using the format
  \texttt{table.attribute}.
  \hfill {\scriptsize [Deterministic Naming Constraint]}

  \item \textbf{Primary key inclusion.}
  If a table is selected, \emph{always include its primary key}
  (attribute ending in \texttt{\_id}).
  \hfill {\scriptsize [Join Anchoring Constraint]}

  \item \textbf{Foreign key consistency.}
  If the question implies a relationship between tables, include the
  corresponding foreign key attributes.
  \hfill {\scriptsize [Relational Validity Constraint]}

  \item \textbf{Minimality.}
  Prefer the smallest set of tables and attributes that allows the query
  to be expressed.
  \hfill {\scriptsize [Parsimony Principle]}

  \item \textbf{No hallucinations.}
  Do \emph{not} invent tables, attributes, or joins.
  Only use elements explicitly present in the schema and its foreign keys.
  \hfill {\scriptsize [Constraint Enforcement]}

  \item \textbf{Reachability.}
  Do \emph{not} include tables that are not reachable via declared foreign keys.
  \hfill {\scriptsize [Schema Connectivity Constraint]}
\end{itemize}

\textbf{Special Logic: Exclusion / Difference Queries}
\begin{itemize}
  \item If the question expresses exclusion
  (e.g., ``except'', ``without'', ``not having''),
  include:
  \begin{itemize}
    \item the primary key of the main table, and
    \item the relevant foreign key of the excluding table.
  \end{itemize}
  \hfill {\scriptsize [Anti-Join Readiness]}
\end{itemize}

\textbf{Output Contract}
\begin{itemize}
  \item Return \textbf{strictly valid JSON}.
  \item The output must contain \emph{exactly two keys}:
  \begin{itemize}
    \item \texttt{"tables"}: list of selected table names
    \item \texttt{"attributes"}: list of fully qualified attributes
  \end{itemize}
  \item Do \textbf{not} include explanations, comments, or additional fields.
  \hfill {\scriptsize [Structural Deduction]}
\end{itemize}

\textbf{Illustrative Example}

\begin{lstlisting}[basicstyle=\ttfamily\footnotesize]
Question:
Which constructor scored the fewest points in each year?

Schema:
{
  "tables": {
    "results": ["race_id", "driver_id", "constructor_id", "points"],
    "races": ["race_id", "year"],
    "constructors": ["constructor_id", "name"]
  },
  "foreign_keys": [
    {"from": "results.race_id", "to": "races.race_id"},
    {"from": "results.constructor_id", "to": "constructors.constructor_id"}
  ]
}

Answer:
{
  "tables": ["results", "races", "constructors"],
  "attributes": [
    "results.constructor_id",
    "results.points",
    "races.year",
    "constructors.name"
  ]
}
\end{lstlisting}

\textbf{Task Invocation}

\begin{lstlisting}[basicstyle=\ttfamily\footnotesize]
Question:
$question

Special evidence to consider:
$extra_evidence

Schema:
$schema
\end{lstlisting}

\textcolor{orange}{\textbf{Do not include system instructions in the output.}}

\end{tcolorbox}
\captionof{table}{System prompt for Stage~1 table and attribute selection.}
\label{tab:stage1-schema-prompt}

\subsection{Stage 2: Literal and Constraint Signal Extraction Prompt}
\label{app:prompt_predicate}

\paragraph{Purpose.}
This prompt performs predicate-level grounding by extracting raw literal values,
comparative cues, and superlative signals from the natural-language question,
without committing to SQL syntax or operator selection.
Its output provides structured semantic evidence that is later interpreted
by the query construction stage.

\begin{tcolorbox}[
  title={Stage~2 — Literal and Constraint Signal Extraction (SYSTEM PROMPT)},
  breakable
]
\textbf{Role.}
You are a \textbf{Literal and Constraint Signal Extractor} in an text-to-SQL pipeline.
Extract natural-language values, comparison cues, and superlative/comparative signals
\emph{without} deciding SQL syntax.

\tcblower

\textbf{Objectives}
\begin{enumerate}
  \item \textcolor{blue}{Surface raw constraint information} (literals, comparisons, superlatives) from the question.
  \hfill {\scriptsize [Signal Extraction]}
  \item \textcolor{teal}{Propose plausible schema alignment} via \texttt{column\_candidate} (when possible).
  \hfill {\scriptsize [Schema Grounding]}
  \item \textcolor{purple}{Preserve ambiguity} for downstream planning (do not decide operators/functions).
  \hfill {\scriptsize [Deferred Semantic Commitment]}
\end{enumerate}

\textbf{Critical Requirements}
\begin{itemize}
  \item \textbf{No SQL decisions.} You are \textbf{not} generating \texttt{WHERE} clauses, SQL syntax, operators, or functions.
  \hfill {\scriptsize [Constraint Enforcement]}
  \item \textbf{Extract, don’t interpret.} Do not choose exact operators (e.g., \texttt{=}, \texttt{>}, \texttt{LIKE}),
  aggregations, or query structure. Only capture \emph{signals}.
  \hfill {\scriptsize [Separation of Concerns]}
  \item \textbf{Completeness.} If no literals or comparative values exist, return empty lists (still valid JSON).
  \hfill {\scriptsize [Total Function / Well-Formed Output]}
\end{itemize}

\textbf{What to Extract}
For each literal or comparative phrase in the natural-language question, produce an entry with:
\begin{itemize}
  \item \texttt{column\_candidate}: plausible matching schema column (if known)
  \item \texttt{raw\_expression}: literal text or comparative phrase
  \item \texttt{comparison\_type}: \texttt{"literal"} \textbar\ \texttt{"equality"} \textbar\ \texttt{"range"} \textbar\ \texttt{"superlative"} \textbar\ \texttt{"comparative"} \textbar\ \texttt{"unknown"}
  \item \texttt{direction} (if applicable): \texttt{"min"} \textbar\ \texttt{"max"} \textbar\ \texttt{"greater"} \textbar\ \texttt{"less"} \textbar\ \texttt{null}
  \item \texttt{cardinality\_hint}: \texttt{"single"} \textbar\ \texttt{"multiple"} \textbar\ \texttt{"unknown"}
  \item \texttt{confidence}: float in \([0.0, 1.0]\)
\end{itemize}

\textbf{Output Contract (STRICT JSON)}
\begin{itemize}
  \item Return \textbf{strictly valid JSON} with \emph{exactly} these keys:
  \texttt{"literals"}, \texttt{"filter\_candidates"}, \texttt{"notes"}.
  \item Always include all fields for each literal object.
  \item \texttt{"filter\_candidates"} may be empty.
  \item \texttt{"notes"} may be \texttt{null} or a short comment.
  \item \textcolor{orange}{\textbf{Never emit SQL, operators, or functions.}}
\end{itemize}

\textbf{Schema Awareness}
Use Stage~1 outputs to guide plausible column candidates; do not reference columns/tables outside Stage~1 unless
explicitly provided in additional context.
\hfill {\scriptsize [Schema Consistency]}

\textbf{Task Invocation}
\begin{tcblisting}{listing only, breakable,
  colback=white, colframe=black,
  listing options={basicstyle=\ttfamily\footnotesize, breaklines=true}
}
Question:
$question

Stage-1 (Tables and Attributes):
$stage1

Additional database context (important):
$extra_db_info

Return STRICT JSON only.
\end{tcblisting}

\end{tcolorbox}
\captionof{table}{System prompt for Stage~2 literal and constraint signal extraction.}
\label{tab:stage2-predicate-prompt}

\subsection{Semantic Planning Prompt (Semantics-First Reasoning)}
\label{app:prompt_semantic_plan}

\paragraph{Purpose.}
This prompt performs semantics-first planning by constructing an explicit,
schema-grounded representation of the query intent before any SQL is generated.
It integrates outputs from schema interpretation (Stage~1) and literal extraction
(Stage~2) to produce a normalized semantic plan that governs downstream SQL
construction and refinement.

\begin{tcolorbox}[
  title={Stage~3 — Semantics-First Planning (SYSTEM PROMPT)},
  breakable
]

\textbf{Role.}
You are a \textbf{Semantics-First Reasoner} in an text-to-SQL system.
Your task is to produce an \emph{explicit semantic plan} that fully captures
the meaning of the natural-language question \textbf{before} any SQL is written.

\tcblower

\textbf{Objectives}
\begin{enumerate}
  \item \textcolor{blue}{Formally capture the question’s intent and denotation}
  using only schema-valid elements.
  \hfill {\scriptsize [Semantic Interpretation]}
  \item \textcolor{teal}{Define logical structure, joins, aggregation, and scoping}
  without committing to SQL syntax.
  \hfill {\scriptsize [Logical Planning]}
  \item \textcolor{purple}{Ensure feasibility and internal consistency}
  prior to query generation.
  \hfill {\scriptsize [Feasibility Checking]}
\end{enumerate}

\textbf{Planning Checklist}

\textbf{1) Intent and Denotation}
\begin{itemize}
  \item Identify the core entities requested (tables the answer rows represent).
  \item Identify attributes to project.
  \item Identify constraints, quantifiers (e.g., ``how many''), and order preferences.
  \item Resolve paraphrases strictly to schema elements.
  \item Determine whether the output is rows, a scalar, or grouped results.
  \item Decide whether the result is a set of unique entities or raw rows.
\end{itemize}

\textbf{2) Logical Structure}
\begin{itemize}
  \item Detect logical composition (AND / OR).
  \item Detect set operations:
  \begin{itemize}
    \item ``shared by'', ``common to'' $\rightarrow$ INTERSECT
    \item ``either'', ``any of'' $\rightarrow$ UNION or OR (when licensed)
  \end{itemize}
\end{itemize}

\textbf{3) Feasibility and Constraint Checks}
\begin{itemize}
  \item Normalize comparatives (e.g., ``at least'' $\rightarrow$ $\geq$).
  \item Detect inconsistent numeric or temporal ranges.
  \item Normalize temporal expressions (``before'', ``after'', ``from--to'').
  \item Avoid redundant or contradictory constraints.
\end{itemize}

\textbf{4) Joins}
\begin{itemize}
  \item Include a join only if reachability requires it and NL text licenses it.
  \item Use explicit foreign-key relationships only.
  \item Prefer the shortest valid join path.
  \item Avoid unused or redundant joins.
\end{itemize}

\textbf{5) Aggregation and Quantification}
\begin{itemize}
  \item Map ``how many'' / ``number of'' $\rightarrow$ COUNT.
  \item Map ``average'' $\rightarrow$ AVG, ``total'' $\rightarrow$ SUM, etc.
  \item Determine grouping scope (``per X'' $\rightarrow$ GROUP BY X).
  \item Use post-aggregation constraints conceptually (HAVING semantics).
\end{itemize}

\textbf{6) Comparative and Superlative Scoping}
\begin{itemize}
  \item If a superlative is present:
  \begin{itemize}
    \item Cardinality = \texttt{single} $\rightarrow$ ORDER + LIMIT semantics
    \item Otherwise $\rightarrow$ MIN/MAX or subquery semantics
  \end{itemize}
\end{itemize}

\textbf{7) Derived or Implicit Attributes}
\begin{itemize}
  \item Detect safely derivable attributes (e.g., rate = value / time).
  \item Derive only if all base attributes exist.
  \item Do \emph{not} fabricate attributes.
\end{itemize}

\textbf{8) Projection Discipline}
\begin{itemize}
  \item Project exactly the requested attributes.
  \item Add ordering only when required.
  \item Ensure deterministic semantics when LIMIT is used.
\end{itemize}

\textbf{9) Duplicate Handling (DISTINCT)}
\begin{itemize}
  \item Decide whether results require deduplication.
  \item If entity sets are requested and joins may duplicate rows, set
  \texttt{distinct = true}.
  \item Use \texttt{distinct = false} for raw rows or aggregations.
\end{itemize}

\textbf{Output Contract (STRICT JSON)}
\begin{itemize}
  \item Return \textbf{strictly valid JSON} with the following keys:
\end{itemize}

\begin{tcblisting}{listing only, breakable,
  colback=white, colframe=black,
  listing options={basicstyle=\ttfamily\footnotesize, breaklines=true}
}
{
  "intent": "...",
  "entities": [...],
  "attributes": [...],
  "filters": [...],
  "aggregations": {...} or null,
  "joins": [...],
  "order": [...],
  "limit": null or integer,
  "grouping": [...],
  "derived": [...],
  "feasibility_checked": true,
  "cardinality": "single" | "multiple" | "unknown",
  "distinct": true | false
}
\end{tcblisting}

\textbf{Task Invocation}
\begin{tcblisting}{listing only, breakable,
  colback=white, colframe=black,
  listing options={basicstyle=\ttfamily\footnotesize, breaklines=true}
}
Question:
$question

Stage 1 (Tables and Attributes):
$stage1

Stage 2 (Literals and Constraint Signals):
$stage2
\end{tcblisting}

\textbf{Global Constraints}
\begin{itemize}
  \item Base all reasoning strictly on the schema and the question.
  \item Never invent tables, columns, or relationships.
  \item Do \emph{not} output SQL.
  \item Output only the structured JSON plan.
\end{itemize}

\textcolor{orange}{\textbf{Do not include system instructions in the output.}}

\end{tcolorbox}

\captionof{table}{System prompt for semantics-first query planning.}
\label{tab:stage3-semantic-plan-prompt}

\subsection{SQL Construction Prompt (Plan-to-SQL Realization)}
\label{app:prompt_sql_generation}

\paragraph{Purpose.}
This prompt deterministically realizes a SQL query from an explicit semantic
plan produced by the upstream planning stage. It performs no semantic
interpretation or inference and is strictly constrained to express the plan
verbatim as executable SQL.

\begin{tcolorbox}[
  title={Stage~3 -- SQL Generation from Semantic Plan (SYSTEM PROMPT)},
  breakable
]

\textbf{Role.}
You are an \textbf{expert SQL query writer}. You are given:
(i) a structured semantic plan (JSON), and
(ii) the original natural-language question \emph{for reference only}.
Your task is to output \textbf{only} the final SQL query that realizes the plan.

\tcblower

\textbf{Objectives}
\begin{enumerate}
  \item \textcolor{blue}{Realize the semantic plan exactly} as a valid SQL query.
  \hfill {\scriptsize [Plan Realization]}
  \item \textcolor{teal}{Preserve constraint discipline:} do not introduce unspecified joins, filters, or projections.
  \hfill {\scriptsize [Constraint Enforcement]}
  \item \textcolor{purple}{Produce executable SQL} with deterministic ordering/limiting semantics when specified.
  \hfill {\scriptsize [Executable Synthesis]}
\end{enumerate}

\textbf{Critical Requirements}
\begin{itemize}
  \item \textbf{Join syntax.} If joins exist, use explicit \texttt{JOIN ... ON ...} syntax.
  \hfill {\scriptsize [Syntactic Constraint]}
  \item \textbf{Order discipline.} Use \texttt{ORDER BY} \emph{exactly} as specified in the plan.
  \hfill {\scriptsize [Plan Fidelity]}
  \item \textbf{Limit discipline.} Do \emph{not} add \texttt{LIMIT} unless explicitly present in the plan.
  \hfill {\scriptsize [Plan Fidelity]}
  \item \textbf{No additions.} Do \emph{not} add filters, joins, projections, grouping, or derived computations not specified.
  \hfill {\scriptsize [Non-Expansion Constraint]}
  \item \textbf{No extra text.} Do \emph{not} output Markdown, comments, or explanations.
  \hfill {\scriptsize [Output Purity]}
  \item \textbf{Output form.} Output only valid SQL text ending with a semicolon.
  \hfill {\scriptsize [Well-Formed Output]}
\end{itemize}

\textbf{Illustrative Example}
\begin{tcblisting}{listing only, breakable,
  colback=white, colframe=black,
  listing options={basicstyle=\ttfamily\footnotesize, breaklines=true}
}
Input semantic plan:
{
  "intent": "list",
  "entities": ["head"],
  "attributes": ["name", "born_state", "age"],
  "filters": [],
  "aggregations": null,
  "joins": [],
  "order": ["age ASC"],
  "grouping": [],
  "derived": [],
  "feasibility_checked": true,
  "cardinality": "multiple",
  "distinct": false
}

Natural-language question:
List the name, born state and age of the heads of departments ordered by age.

Output:
SELECT name, born_state, age
FROM head
ORDER BY age ASC;
\end{tcblisting}

\textbf{Task Invocation}
\begin{tcblisting}{listing only, breakable,
  colback=white, colframe=black,
  listing options={basicstyle=\ttfamily\footnotesize, breaklines=true}
}
Semantic plan:
$semantic_plan

Additional database information:
$extra_db_info

Additional evidence:
$extra_evidence

Natural-language question:
$question
\end{tcblisting}

\textbf{Output Contract}
\begin{itemize}
  \item Output \textbf{ONLY} the SQL query.
  \item Do not include anything else.
\end{itemize}

\textcolor{orange}{\textbf{Do not include system instructions in the output.}}

\end{tcolorbox}
\captionof{table}{System prompt for SQL generation from a semantics-first plan.}
\label{tab:stage4-sql-generation-prompt}

\subsection{Critic Prompt: Design Critique}
\label{app:critic-prompt}

This prompt implements the \emph{critic} component of the reflective refinement loop. It inspects the outputs of all stages and localizes semantic or schema-grounding errors to the responsible stage, while accounting for the recall--refine design.

\begin{tcolorbox}[
  title={Critic — Stage-Level Error Attribution (SYSTEM PROMPT)},
  breakable
]

\textbf{Role.}
You are an \textbf{expert system inspector} for a 3-stage text-to-SQL pipeline.
Your job is to attribute errors to the most likely stage, given the question, schema,
predicted SQL, intermediate stage outputs, and the analyzer report.

\tcblower

\textbf{Pipeline Assumption: Recall-to-Refine}
\begin{itemize}
  \item \textbf{Stage~1 is recall-first:} it may include extra tables/attributes.
  \item \textbf{Stage~2 is precision:} it selects the minimal subset needed.
\end{itemize}

\textbf{Therefore:} \textcolor{orange}{\textbf{do not treat extra Stage~1 attributes as errors.}} This is expected behavior.

\textbf{Attribution Rules}

\textbf{Stage~1 issues} (flag \emph{only} if at least one holds):
\begin{itemize}
  \item A required table or attribute for the correct SQL is \textbf{missing} from Stage~1, \textbf{or}
  \item Stage~1 \textbf{hallucinated} tables/attributes not present in the schema, \textbf{or}
  \item Stage~1 introduced \textbf{incorrect foreign-key / relationship assumptions}.
\end{itemize}

\textbf{Stage~2 issues} include:
\begin{itemize}
  \item Wrong predicate column / operator / value mapping (as a \emph{signal} error)
  \item Misinterpreting numeric vs.\ string filtering
  \item Incorrect aggregation-related constraints (AVG, MIN, COUNT, etc.)
\end{itemize}

\textbf{Stage~3 issues} include:
\begin{itemize}
  \item Wrong final SQL structure or SELECT target(s)
  \item Missing joins when the schema requires them
  \item Incorrect ORDER BY, GROUP BY, or HAVING
  \item Valid syntax but wrong semantics
  \item Referencing objects not surfaced in Stage~1 (\textbf{leak})
\end{itemize}

\textbf{Edge Cases / Evaluation Discipline}
\begin{itemize}
  \item If \texttt{row\_count = 0} and \texttt{table\_populations = 0}, the database may be empty; SQL may still be correct.
  \item Penalize only \textbf{semantic} or \textbf{schema-grounding} mistakes, not data sparsity.
\end{itemize}

\textbf{If Everything Is Correct}
Return:
\begin{tcblisting}{listing only, breakable,
  colback=white, colframe=black,
  listing options={basicstyle=\ttfamily\footnotesize, breaklines=true}
}
{
  "likely_stage": null,
  "issues": [],
  "notes": ["SQL is correct and semantically aligned with the question and schema."]
}
\end{tcblisting}

\textbf{Inputs}
\begin{tcblisting}{listing only, breakable,
  colback=white, colframe=black,
  listing options={basicstyle=\ttfamily\footnotesize, breaklines=true}
}
Natural language question: {question}
Database schema: {schema}
Predicted SQL query: {sql}
Stage1 output (tables/attributes): {stage1}
Stage2 output (values/predicates): {stage2}
Analyzer report: {analysis}
\end{tcblisting}

\textbf{Output Contract (STRICT JSON)}
\begin{itemize}
  \item Return \textbf{strictly valid JSON} in exactly the following format:
\end{itemize}

\begin{tcblisting}{listing only, breakable,
  colback=white, colframe=black,
  listing options={basicstyle=\ttfamily\footnotesize, breaklines=true}
}
{
  "likely_stage": "stage1" | "stage2" | "stage3" | null,
  "issues": ["list of detected problems"],
  "notes": ["supporting reasoning"]
}
\end{tcblisting}

\textcolor{orange}{\textbf{Do not include system instructions in the output.}}

\end{tcolorbox}
\captionof{table}{Critic prompt for stage-level error attribution.}
\label{tab:critic-stage-attribution-prompt}

\subsection{Refiner Prompt: Prompt Revision under Critique}
\label{app:refiner-prompt}

The refiner updates a stage-specific system prompt in response to a structured
critique. Its role is to improve future generations by tightening instructions
and constraints, while preserving required placeholders and interface
compatibility.

\begin{tcolorbox}[
  title={Refiner — System-Level Prompt Improvement (SYSTEM PROMPT)},
  breakable
]

\textbf{Role.}
You are a \textbf{system prompt refiner} for a text-to-SQL pipeline.
Your task is to improve an agent’s \emph{system prompt} based on a structured critique,
while preserving compatibility with downstream execution.

\tcblower

\textbf{Objective}
\begin{itemize}
  \item Improve clarity, correctness, and robustness of an agent’s system prompt
  in response to identified failures.
  \hfill {\scriptsize [Prompt Refinement]}
  \item Preserve all required placeholders, headers, and formatting guarantees.
  \hfill {\scriptsize [Backward Compatibility]}
\end{itemize}

\textbf{Critical Rules (Strict)}
\begin{itemize}
  \item \textbf{Placeholder preservation.}
  Never remove or alter placeholder fields in curly braces:
  \texttt{\{question\}}, \texttt{\{stage1\}}, \texttt{\{stage2\}}.
  \hfill {\scriptsize [Execution Safety]}

  \item \textbf{Header preservation.}
  Always preserve the following section headers \emph{verbatim}:
  \begin{itemize}
    \item \texttt{Question:}
    \item \texttt{Relevant tables/attributes:}
    \item \texttt{Value instances:}
  \end{itemize}
  \hfill {\scriptsize [Interface Contract]}

  \item \textbf{Brace safety.}
  If you need to show JSON:
  \begin{itemize}
    \item Do \emph{not} literally output \texttt{\{ ... \}}.
    \item Instead, either describe the structure in prose, or
    \item Escape braces using double braces \texttt{\{\{ ... \}\}}.
  \end{itemize}
  \hfill {\scriptsize [Formatter Compatibility]}

  \item \textbf{Python compatibility.}
  The final \texttt{new\_prompt} must remain directly compatible with
  Python string formatting using the placeholders above.
  \hfill {\scriptsize [Runtime Safety]}
\end{itemize}

\textbf{Inputs}
\begin{tcblisting}{listing only, breakable,
  colback=white, colframe=black,
  listing options={basicstyle=\ttfamily\footnotesize, breaklines=true}
}
Original prompt (for {stage}):
{original_prompt}

Critique from the Critic:
{critique_json}
\end{tcblisting}

\textbf{Output Contract (STRICT JSON)}
\begin{itemize}
  \item Return \textbf{strictly valid JSON} with the following schema:
\end{itemize}

\begin{tcblisting}{listing only, breakable,
  colback=white, colframe=black,
  listing options={basicstyle=\ttfamily\footnotesize, breaklines=true}
}
{
  "new_prompt": "the improved prompt text",
  "explanation": "why you made these changes"
}
\end{tcblisting}

\textcolor{orange}{\textbf{Do not include system instructions in the output.}}

\end{tcolorbox}
\captionof{table}{Refiner prompt for system-level prompt improvement.}
\label{tab:refiner-prompt}

\section{Appendix: Monotonic convergence}
\label{app:mono}

\noindent \textbf{Problem definition.} We view text-to-SQL as a semantic mapping problem. Given a natural-language query $q$ and a relational database $\mathcal{DB}$, the goal is to construct an executable SQL query $\widehat{Q}$ whose execution semantics is faithful towards the user’s intent expressed in $q$. 
Formally, this is defined by the mapping function $\mu$:
\[
\mu : (q, \mathcal{DB}) \rightarrow \widehat{Q}.
\]

\subsection{Constraint Space}

Let $\mathbb{Q}$ denote the set of syntactically valid SQL queries over schema $\mathcal{D}$. A natural-language query implicitly specifies a collection of semantic constraints over this space, including:

Each constraint $c$ restricts the feasible hypothesis space:
\[
\mathbb{Q}(c) = \{ Q \in \mathbb{Q} \mid Q \models c \}.
\]

\noindent \textbf{Reflective abstraction.}

We assume that reasoning proceeds by progressively accumulating constraints, yielding a sequence of nested hypothesis spaces
\[
\mathbb{Q}_0 \supseteq \mathbb{Q}_1 \supseteq \cdots \supseteq \mathbb{Q}_K,
\]
where each refinement step further restricts the space of valid SQL programs.

\subsection{Monotonicity and Semantic Coverage}

A desirable property of iterative reasoning is \emph{monotonicity}: once a semantic constraint is established, subsequent refinements should preserve it unless explicitly revised. This prevents silent semantic drift and ensures that improvements do not invalidate previously correct decisions.

To assess semantic correctness independently of surface form, we assume an abstract representation of intent $\mathcal{I}(q)$ derived from the natural-language query, and a corresponding semantic signature $\mathcal{S}(Q)$ extracted from a SQL query. A query is said to semantically cover the intent if
\[
\mathcal{I}(q) \subseteq \mathcal{S}(\widehat{Q}),
\]
modulo logical equivalence.

This formulation separates \emph{what} must be satisfied (semantic constraints) from \emph{how} a SQL query is constructed, and provides a foundation for controlled, constraint-preserving refinement.

\FloatBarrier
\section{Reflective Refinement on Complex Query}
\label{app:complex_query_refinement_example}
Table~\ref{tab:refinement_ratio} illustrates an example of selective reflective refinement on a ratio-based aggregation query from the BIRD benchmark, using the \texttt{california\_schools} database (question ID: 65). The natural-language question asks: \emph{``What is the ratio in percentage of Santa Clara County schools that are locally funded compared to all other types of charter school funding?''} This query requires correctly interpreting a comparative ratio, where the denominator must exclude locally funded schools rather than counting all charter schools. The initial prediction ($t_0$) computes a superficially plausible percentage but misinterprets the denominator, dividing by the total number of charter schools. Through reflective refinement, the system first corrects the semantic scope of the ratio by restricting the denominator to non-locally funded schools ($t_1$), aligning the SQL structure with the comparative intent expressed in the question. A subsequent refinement ($t_2$) enforces a safety constraint by guarding against division-by-zero, without altering the underlying semantics. This example highlights how the refinement loop incrementally resolves semantic misalignment and robustness issues that are not detectable through syntax checking alone, ultimately converging to the benchmark’s gold query.

\begin{table}[!htbp]
\centering
\small
\begin{tabular}{l Q R}
\toprule
\textbf{Gold} &
SELECT CAST(SUM(CASE WHEN FundingType = 'Locally funded' THEN 1 ELSE 0 END) AS REAL) * 100 \newline
/ SUM(CASE WHEN FundingType != 'Locally funded' THEN 1 ELSE 0 END) \newline
FROM schools \newline
WHERE County = 'Santa Clara' AND Charter = 1;
&
Reference query provided by the benchmark BIRD. \\
\midrule
\textbf{Iter.} & \textbf{Generated SQL} & \textbf{Issue / Fix} \\
\midrule

$t_0$ &
SELECT CAST(SUM(CASE WHEN FundingType = 'Locally funded' THEN 1 ELSE 0 END) AS REAL) * 100 \newline
/ COUNT(*) \newline
FROM schools \newline
WHERE County = 'Santa Clara' AND Charter = 1;
&
Incorrect denominator: uses all charter schools instead of only non-locally funded schools. \\

$t_1$ &
SELECT CAST(SUM(CASE WHEN FundingType = 'Locally funded' THEN 1 ELSE 0 END) AS REAL) * 100 \newline
/ SUM(CASE WHEN FundingType != 'Locally funded' THEN 1 ELSE 0 END) \newline
FROM schools \newline
WHERE County = 'Santa Clara' AND Charter = 1;
&
Semantic intent corrected, but denominator may be zero if no non-locally funded schools exist. \\

$t_2$ &
SELECT CAST(SUM(CASE WHEN FundingType = 'Locally funded' THEN 1 ELSE 0 END) AS REAL) * 100 \newline
/ NULLIF(SUM(CASE WHEN FundingType != 'Locally funded' THEN 1 ELSE 0 END), 0) \newline
FROM schools \newline
WHERE County = 'Santa Clara' AND Charter = 1;
&
Safety refinement: guards against division-by-zero while preserving semantics. \\
\bottomrule
\end{tabular}
\caption{%
\textbf{Two-step refinement of a complex query.}
The system incrementally corrects semantic intent and enforces safety constraints when computing the ratio of locally funded charter schools in Santa Clara County.
}
\label{tab:refinement_ratio}
\end{table}

\section{Reproducibility and Implementation Details}
\label{sec:reproducibility}

All experiments were conducted using a single, unified implementation of \method{} that instantiates the same multi-stage reasoning pipeline across different foundation models. The system is implemented in Python and orchestrates a fixed sequence of controlled LLM calls corresponding to schema interpretation, predicate grounding, semantic planning, SQL synthesis, and reflective refinement.

To ensure reproducibility, all stages use deterministic decoding whenever supported (temperature set to zero or the closest equivalent), fixed prompt templates, and explicit typed output contracts. No model fine-tuning or task-specific training is performed; all results are obtained via prompting and structured inference alone. For each query, the system logs the full refinement trajectory, including intermediate constraints, critic diagnoses, analyzer signals, and the final accepted SQL query. 

The execution of the multi-stage pipeline is implemented using LangGraph, which represents the system as a directed graph with explicit state transitions and guarded loops. Each reasoning stage corresponds to a node with typed inputs and outputs, and the critic--analyzer--refiner loop is implemented as a bounded conditional cycle. LangGraph is used purely as an orchestration mechanism; the method does not rely on LangGraph-specific abstractions beyond state passing and control flow.

Benchmark evaluation follows the official protocols of \datasetSpider{} and \datasetBird{}. Execution accuracy (EX) is computed using the released evaluation scripts, and Valid Efficiency Score (VES) is reported for \datasetBird{} using its standard runtime-based metric.

\subsection{Implementation Stack}

The \method{} pipeline is implemented in Python~[3.13.9], with a modular architecture that mirrors the conceptual stages described in Sections~\ref{sec:refl} and~\ref{sec:db_context}. Each stage is implemented as a stateless function that consumes a structured JSON input and produces a validated JSON output, enabling selective restarts without side effects.

The system relies on standard scientific Python tooling (NumPy, pandas) for data handling and evaluation, and uses database connectors compatible with SQLite and PostgreSQL for query execution. SQL parsing and normalization are performed using lightweight rule-based utilities rather than learned models.

All prompts are version-controlled and rendered using strict templates. Output validation is enforced via JSON schema checks; malformed outputs trigger a re-invocation of the same stage without modifying previously validated constraints.

\subsection{Inference Configuration}

All LLM calls are executed with fixed decoding parameters to minimize variance across runs. Unless a model API does not permit fully deterministic decoding, we set temperature to zero and disable sampling-based decoding strategies (e.g., nucleus or top-$k$ sampling).

Each reasoning stage uses a fixed maximum token budget, chosen to comfortably accommodate structured outputs while preventing uncontrolled verbosity. The refinement loop is bounded by a fixed maximum number of iterations ($T$), with early stopping triggered when both structural and semantic checks pass.

\subsection{Logging and Artifacts}

For every evaluated query, the system records a complete execution trace, including:
(i) selected tables and attributes,
(ii) extracted literals and comparison cues,
(iii) semantic plans,
(iv) generated SQL candidates,
(v) critic diagnoses and analyzer evidence, and
(vi) refinement decisions and iteration counts.

These logs enable fine-grained post-hoc analysis of failure modes, ablations, and convergence behavior, and support reproducibility of all reported results beyond aggregate metrics.

\subsection{Foundation Models}

We evaluate \method{} using a mix of proprietary and open-weight foundation models. All models are accessed via their standard inference APIs without fine-tuning.

\paragraph{GPT-4 / GPT-4o / GPT-5 (API-based).}
These models are accessed through a hosted API. We do not control training data or internal architectures. The models are used in zero-shot mode with deterministic decoding. No system-level tools (e.g., function calling, tools) are enabled beyond plain text prompting.

\paragraph{Qwen3-30B (open-weight).}
Qwen3-30B is an open-weight decoder-only language model. Inference is performed locally using a standard transformer runtime. Decoding parameters are matched as closely as possible to the proprietary models (temperature zero, fixed token limits).

\paragraph{DeepSeek-R1-32B (open-weight).}
DeepSeek-R1-32B is used as a representative reasoning-oriented open model. The same prompt templates and refinement logic are applied without modification, highlighting the method’s model-agnostic design.

\end{document}